\documentclass[aps,prr,twocolumn,floatfix,showpacs,nofootinbib,groupedaddress,superscriptaddress]{revtex4-1}
\pdfoutput=1
\usepackage[linkcolor=blue,colorlinks=true,breaklinks=true,citecolor=blue,urlcolor=blue]{hyperref}
\usepackage{graphicx,graphics}  
\usepackage{dcolumn}   
\usepackage{braket}
\usepackage[mathscr]{euscript}
\usepackage{amsbsy}
\usepackage{bm}        
\usepackage{amssymb} 
\usepackage{amsmath}
\usepackage{xcolor}
\usepackage[caption=false,labelformat=simple]{subfig}

\makeatletter
\newcommand{\phantomlabelabovecaption}[2]{
	\protected@write\@auxout{}{
		\string\newlabel{#2}{
			{\number\numexpr\thefigure+1\relax#1}{\thepage}
			{\number\numexpr\thefigure+1\relax#1}{#2}{}
		}
	}
	\hypertarget{#2}{}
}
\makeatother

\begin{document}

\title{Strain induced tunable band gap and optical properties of graphene on hexagonal boron nitride}
\author{Priyanka Sinha}
\email{sinhapriyanka2012@gmail.com}
\affiliation{Department of Physical Sciences, Indian Institute of Science Education and Research Kolkata\\ Mohanpur-741246, West Bengal, India}
\author{Prasanta K. Panigrahi}
\email{pprasanta@iiserkol.ac.in}
\affiliation{Department of Physical Sciences, Indian Institute of Science Education and Research Kolkata\\ Mohanpur-741246, West Bengal, India}
\affiliation{Center for Quantum Science and Technology (CQST), Siksha o Anusandhan (SOA) University, Bhubaneswar-751030, Odisha, India}
\author{Bheemalingam Chittari}
\email{bheemalingam@iiserkol.ac.in}
\affiliation{Department of Physical Sciences, Indian Institute of Science Education and Research Kolkata\\ Mohanpur-741246, West Bengal, India}
\date{\today}
\begin{abstract}
In this study, we highlight the potential of strain engineering in graphene/hBN (hexagonal Boron nitride) 2D heterostructures, enabling their use as wide-range light absorbers with significant implications for optoelectronic applications. We systematically investigate the electronic and optical properties of graphene/hBN under the application of strain, considering various stacking geometries within the framework of density-functional theory (DFT). The semimetallic graphene layer upon aligning on the insulating hexagonal boron nitride sheet opens a few tens of meV band gap at the Dirac point due to the induced on-site energy differences on the two sublattices of graphene. Here, we demonstrate that by simultaneously tuning the interlayer distance and lattice constant, this band gap can be significantly increased to 1 eV. Interestingly, in both scenarios (small and large band gaps), the material undergoes a transition from a semiconductor to a semimetallic state. Importantly, the tunability of this band gap is strongly influenced by the specific stacking configuration. We further explored the optical properties across a broad spectrum, revealing that the presence of a strain-induced band gap fundamentally alters how light interacts with the system.
\end{abstract}
\maketitle

\section{Introduction}\label{I}
In recent years, van der Waals (vdW) heterostructures \cite{geim2}, which are the stacking of two-dimensional (2D) crystals, have attracted extensive research interests owing to their remarkable properties \cite{la,brit,haigh}. Graphene on hexagonal boron nitride (hBN) is an example of such a heterostructure composed of a monolayer graphene and a monolayer hBN coupled by vdW interactions. Both have similar honeycomb structures with a lattice mismatch of less than 2\% \cite{Giovannetti,pakdel}. The experimental lattice constants of graphene and boron nitride are 2.460~\AA~and 2.504~\AA, respectively. Graphene is a semimetal (zero band gap), whereas hBN is an insulator with a wide direct band gap of $\sim$5.97~eV \cite{tani}, as indicated by their electronic behavior. Elucidating the possibilities of the emergence of a band gap in graphene is crucial \cite{pp} and challenging for developing graphene-based devices. Unlike graphene nanoribbons \cite{wa}, chemical functionalization \cite{balog}, etc., the presence of a substrate offers the capability to induce a band gap in graphene without degrading its physical properties. First-generation devices based on graphene and SiO$_{2}$/Si substrates \cite{kang,maas,fan} manifest inadequate electronic transport, primarily due to surface exposure and environmental disorder. Alternatively, researchers have discovered that the hBN substrate is a superior candidate, making graphene a promising component for field-effect transistors. It also features minimal charged impurities and large surface phonon energy and is fabricated experimentally \cite{dean}. The potential lies in the fact that vdW heterostructures can be used in a sophisticated manner to modulate the emergent device properties arising from the interfacial interactions between the stacked materials. Graphene and hBN can be effortlessly combined based on the required stacking geometries, owing to the atomically thin layers without any lattice-matching constraints. Thus, the combination of graphene and hBN heterostructures provides novel prospects for optoelectronic devices \cite{son,band,viti,song} such as light-emitting diodes \cite{with}, photodetectors and autocorrelators \cite{shi}, solar-cells \cite{meng}, and many more with enhanced electron mobilities. 

Moreover, the implementation of strain presents a viable avenue to modify the electronic properties and phonon spectrum of graphene \cite{choi} and hBN. Further, the modulation can be achieved using various approaches such as electric or magnetic field, pressure, doping, etc. The application of strain in vdW heterostructure specifically helps us to achieve an additional level of precision in controlling material properties \cite{j,s}. Usually, the absorption of light for single-layer graphene is 2.3\% over a broad wavelength range and the value increases linearly with the increase of the number of layers \cite{nair}. However, the determination of strongly anisotropic dielectric properties of hBN is a challenging task in the overall optical response. Recent experimental studies have successfully employed imaging spectroscopic ellipsometry, utilizing a simultaneous analysis of the optical response of hBN with graphene monolayers. It has been demonstrated that hBN substrates have the capacity to significantly amplify the absorption in graphene by approximately 60\% over a broad range of spectrum ($\approx$ 250-950 nm) \cite{tok}. A lot of potential applications, for example, the energy storage capacity of a material as well as charge screening in a material depend on its polarizability and dielectric constant. Previously, it has been observed that the optical conductivity of graphene \cite{peres} is tunable within the visible range by rotating the graphene sheet on the hBN substrate \cite{slot}. Nevertheless, many aspects of the electronic and optical properties of graphene on hBN under the application of strain remain largely unexplored.

In this study, we mainly focus on the effects of bi-axial strains on the electronic and optical properties of graphene/hBN heterostructures via first-principles calculations. Initially, we explore the possibility of tuning the electronic properties by changing the interlayer distance between the graphene and the hBN layers and the lattice constant of the system. Due to the emergence of unique electronic properties, it is of fundamental and practical interest to study the influence of hBN on the optical properties of graphene. To see the optical response, we calculate the dielectric constant (both the real and the imaginary parts) and electron energy loss (calculated from the imaginary part of the dielectric function, which measures the absorption spectrum) in the graphene/hBN heterostructures. Next, we consider strain to observe the effects of it on the optical properties of the system. Our findings reveal that the presence of hBN can modify the electronic and optical properties of graphene significantly under the applied strain.\par 
We have organized our paper as follows. In Sec.~(\ref{II}), we describe the first-principles electronic structure calculation method in detail. The formula to calculate the components of the dielectric constant is also described in Sec.~(\ref{II}). Our results have been discussed in Sec.~(\ref{III}). In Sec.~(\ref{IIIA}), we first discuss the ground-state properties, the electronic band structure, and the projected density of states of the system. Then, we present our results for the variation of band gap and the DOS in the presence of strain in Sec.~(\ref{IIIB}). Section~(\ref{IIIC}) is devoted to the study of the optical properties and the effect of strain on the optical properties. Finally, we conclude our results in Sec.~(\ref{IV}).
\begin{figure}[!t]
\begin{center}
\hspace*{-0.6 cm}		
\subfloat{\includegraphics[width=0.4\textwidth]{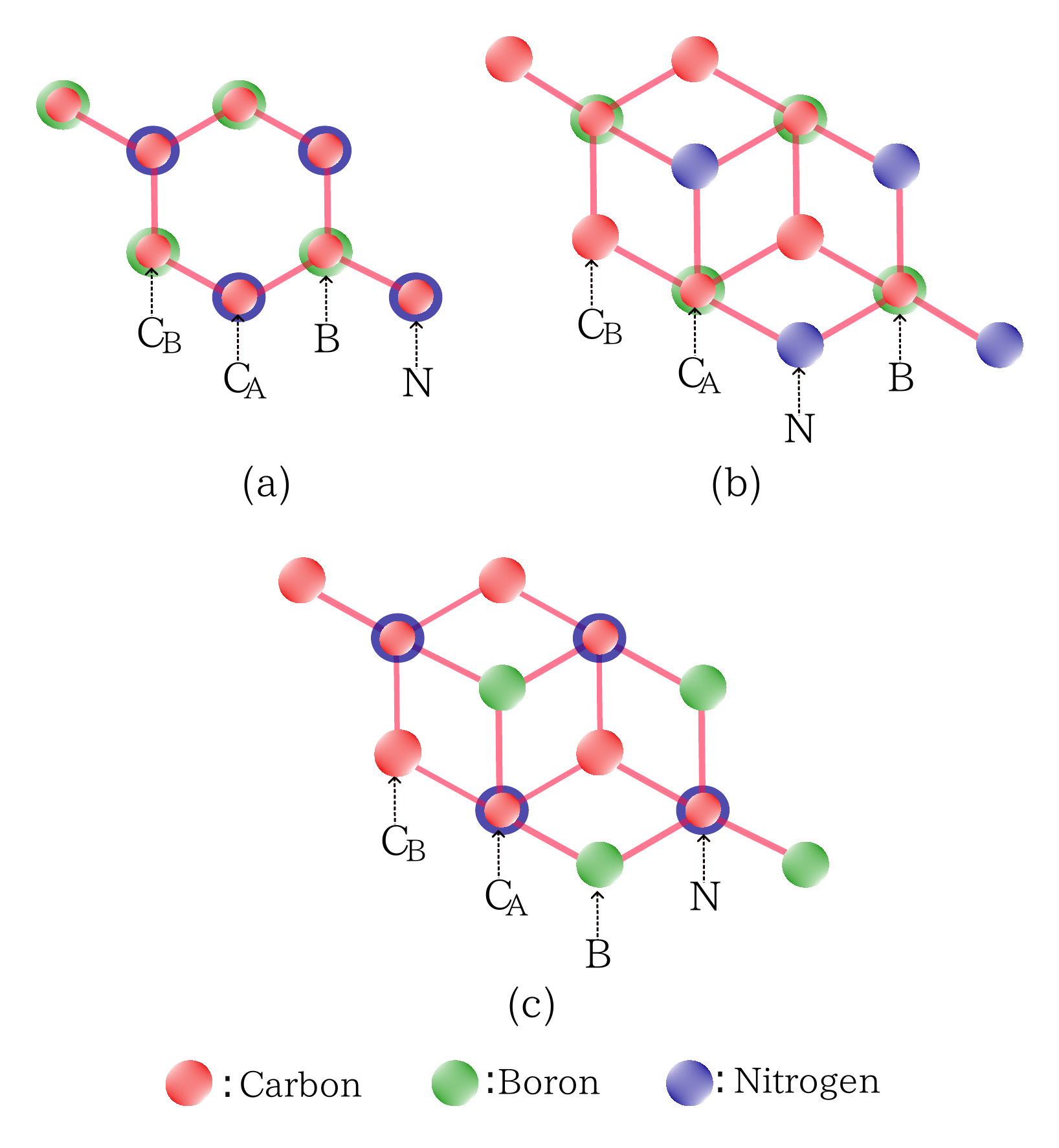}}
\caption{(Color online) A Schematic diagram of three different stacking configurations: (a) AA-stacking, (b) AB-stacking, and (c) BA-stacking of graphene/hBN heterostructure is shown. The symbols C$_{\text{A}}$ and C$_{\text{B}}$ refer to carbon atoms with two distinct sublattices denoted by A and B, while B and N represent Boron and Nitrogen atoms, respectively.}
\label{fig:1}
\end{center}
\end{figure}
\begin{figure}[!t]
	\begin{center}
		\subfloat{\includegraphics[width=0.4\textwidth]{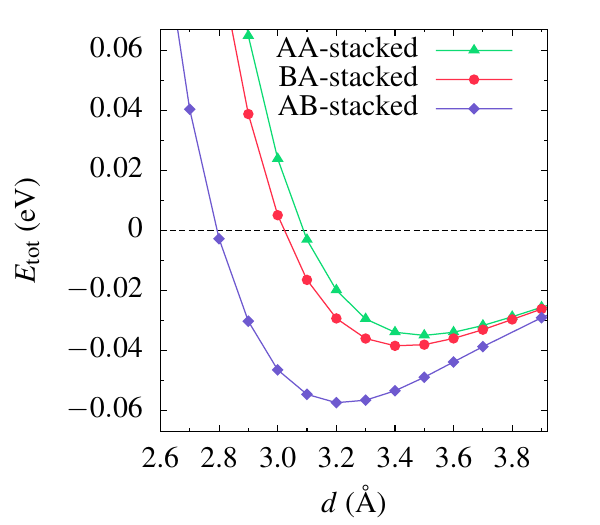}}
		\caption{(Color online) The total energy, $E_{\text{tot}}$ (in units of eV) is plotted as a function of interlayer distance, $d$ (in units of \AA) for AA-stacking (denoted by green curve), BA-stacking (denoted by red curve), and AB-stacking (denoted by blue curve) of graphene/hBN heterostructure.}
		\label{fig:2}
	\end{center}
\end{figure}
\begin{figure}[!t!]
	\begin{center}
		\subfloat{\includegraphics[width=0.4\textwidth]{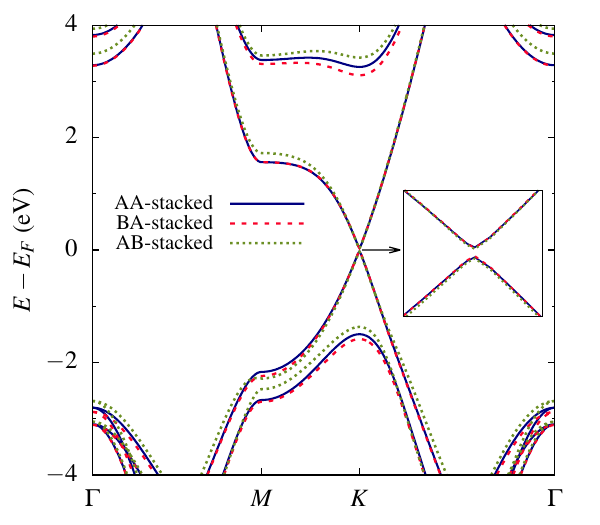}}
		\caption{(Color online) Band structure along the high symmetry point ($\Gamma$ $\rightarrow$ $M$ $\rightarrow$ $K$ $\rightarrow$ $\Gamma$) is shown for all three stackings of graphene/hBN heterostructure. The inset shows the zoomed view near the $K$-point.}
		\label{fig:3}
	\end{center}
\end{figure}
\begin{figure}[!ht!]
	\begin{center} \hspace*{1 cm}
		\subfloat{\includegraphics[width=0.65\textwidth]{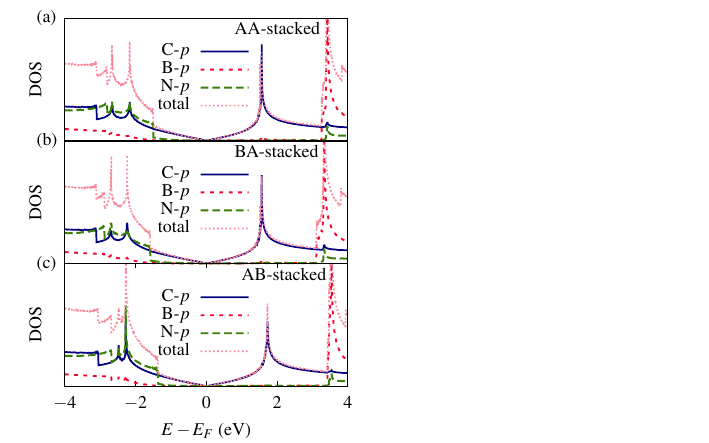}}
		\caption{(Color online) The density of states (DOS) (projected and the total) are shown for (a) AA-stacking, (b) BA-stacking, and (c) AB-stacking of graphene/hBN heterostructure. The projected DOS for carbon (indicated by blue), boron (indicated by red), and nitrogen (indicated by green) are shown with a projection on the p-orbital states (in-plane and out-of-plane).}
		\label{fig:4}
	\end{center}
\end{figure}
\begin{figure}[!t!]
	\begin{center}
		\subfloat[]{\includegraphics[width=0.25\textwidth]{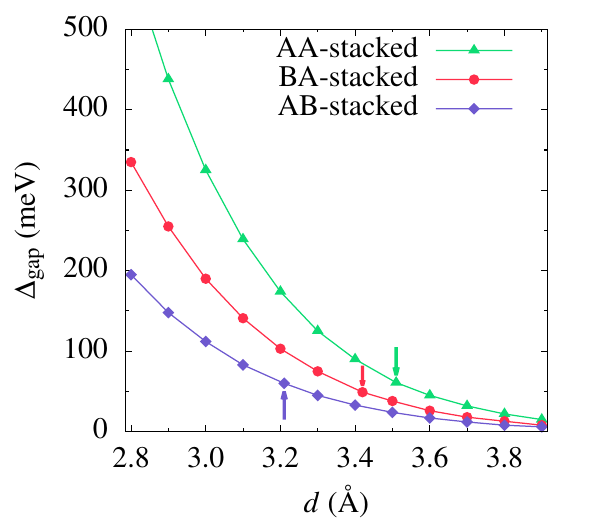}\label{fig:5a}}
		\subfloat[]{\includegraphics[width=0.25\textwidth]{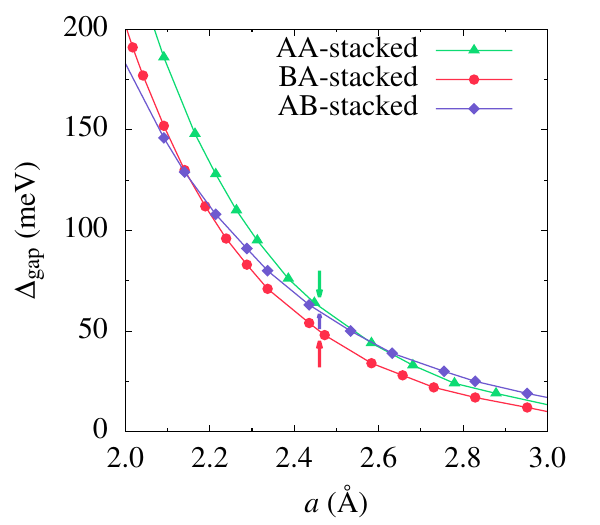}\label{fig:5b}}
		\caption{(Color online) (a) The energy band gap, $\Delta_{\text{gap}}$ (in units of meV) is plotted as a function of interlayer distance, $d$ (in units of \AA) for AA-stacking (denoted by green curve), BA-stacking (denoted by red curve), and AB-stacking (denoted by blue curve). The arrows indicate the value of $\Delta_{\text{gap}}$ at equilibrium separation. (b) The energy band gap, $\Delta_{\text{gap}}$ (in units of meV) is plotted as a function of lattice constant, $a$ (in units of \AA) for AA-stacking (denoted by green curve), BA-stacking (denoted by red curve), and AB-stacking (denoted by blue curve). The arrows indicate the value of $\Delta_{\text{gap}}$ at lattice constant, $a=2.46$ \AA.}
		\label{fig:5}
	\end{center}
\end{figure}
\begin{figure*}[!ht!]
	\begin{center}
		\phantomlabelabovecaption{(a)}{fig:6a}
		\phantomlabelabovecaption{(b)}{fig:6b}
		\phantomlabelabovecaption{(c)}{fig:6c} 
		\subfloat{\includegraphics[width=0.95\textwidth]{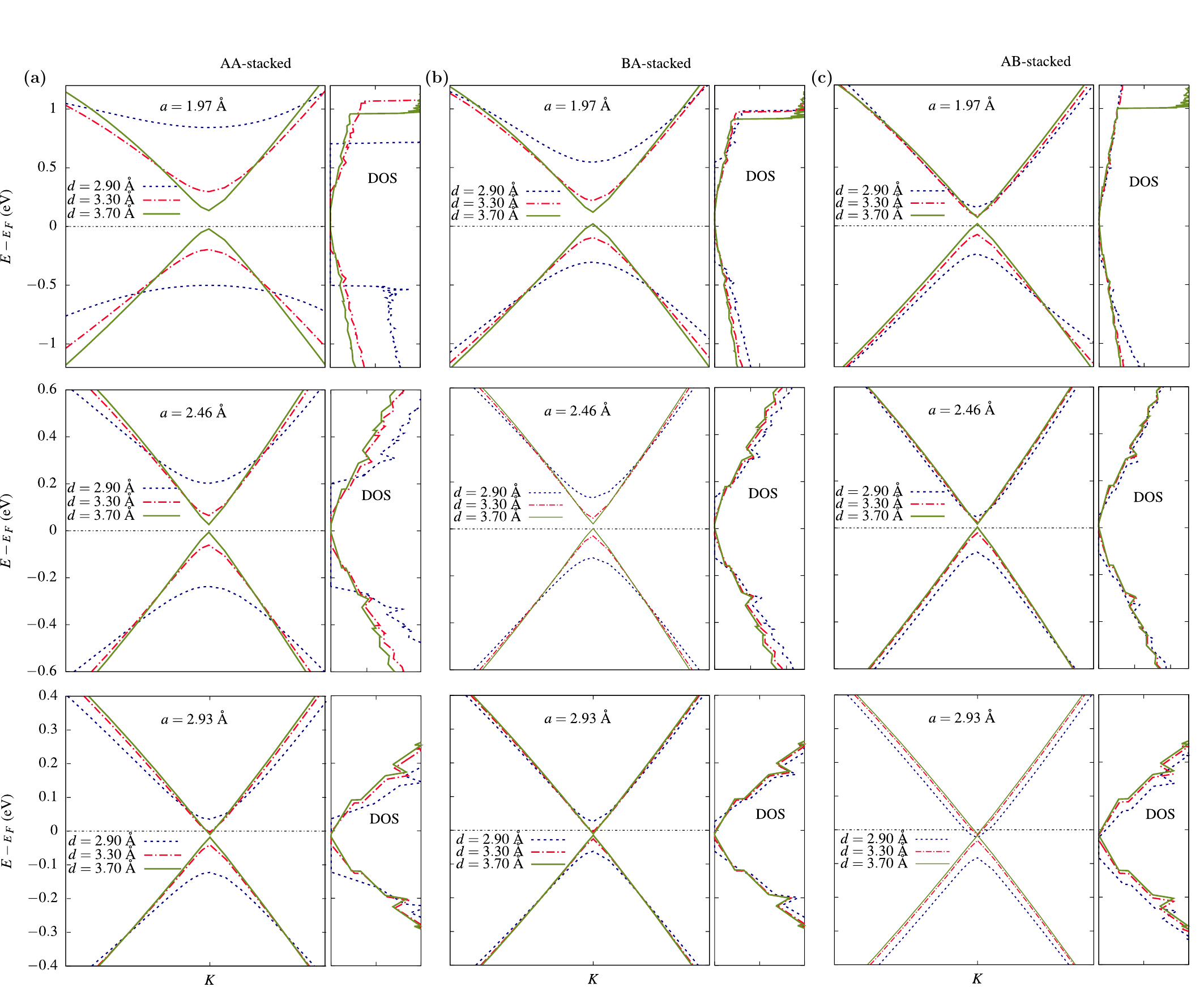}}
		\caption{(Color online) The electronic band structures and the corresponding DOS are shown for different interlayer distance, $d$ (in units of \AA) and lattice constant, $a$ (in units of \AA) for (a) AA-stacking (left panel), (b) BA-stacking (middle panel), and (c) AB-stacking (right panel) of graphene/hBN heterostructure.}
		\label{fig:6}
	\end{center}
\end{figure*}
\section{Methodology}\label{II}
To study the electronic and optical properties of graphene/hBN heterostructures within the DFT framework, a standard unit cell is employed for both graphene and hBN. To avoid the lattice mismatch, the lattice constant of hBN is altered from its experimental value \cite{cate} of 2.504\AA~to 2.46\AA~, which is similar to that of graphene \cite{castro}. In Fig.~\ref{fig:1}, we have shown three different possible configurations of graphene/hBN heterostructure (a) AA-stacking, where one carbon atom C$_{\text{A}}$ is situated directly above the nitrogen (N) atom and the other carbon atom C$_{\text{B}}$ is over the boron (B) atom, (b) AB-stacking, one carbon atom C$_{\text{A}}$ is positioned above B atom, while the other C$_{\text{B}}$ atom is positioned at the center of the hexagonal hollow in hBN, and (c) BA-stacking, where one C$_{\text{A}}$ atom is positioned above N atom, while the other C$_{\text{B}}$ atom is positioned at the center of the hexagonal hollow in hBN. Although the interlayer distances with minimum energy vary among different stackings of graphene/hBN heterostructure \cite{zoll}, it has been demonstrated that the AB stacking configuration represents the most energetically favorable arrangement for the graphene/hBN heterostructure \cite{san,jose,balu}.\par
To perform our DFT calculation, we have used the Quantum Espresso (QE) software package \cite{gia} under local density approximation (LDA) based on the plane-wave density functional theory. All computations were carried out using Perdew-Zunger (pz) pseudopotential approximations, with an energy cutoff of 90 eV to achieve satisfactory convergence. The tetrahedron method was used to calculate the DOS. The convergence criteria defined for energy and force calculations were 10$^{-8}$ Ry and 10$^{-5}$ Ry/Bohr, respectively. A uniform Monkhorst–Pack $k$-point mesh of 96$\times$96$\times$1 is used for our system. A minimum of 25\AA~of vacuum was introduced in the $z$-direction to avoid interactions between the layers. To induce strain in the heterostructure, we modified the lattice constant up to $\pm$20\%.

Further, we performed the calculations for optical properties using the Epsilon package included with Quantum Espresso, based on the random phase approximation (RPA). The optical properties of the system can be represented by its dielectric constant. To model the complex dielectric function $\varepsilon(\omega)$, we have employed a superposition of Lorentz oscillators, expressed as 
\begin{equation}
\varepsilon(\omega) = \varepsilon_{1}(\omega) + i\varepsilon_{2}(\omega),
\end{equation}
where $\varepsilon_{1}(\omega)$ and $\varepsilon_{2}(\omega)$ are the real and imaginary parts of the dielectric constant of the system. The dielectric constant is caused by various kinds of displacement polarization inside the material and represents the energy storage term of the material. The imaginary part of the complex permittivity, $\varepsilon_2(\omega)$ is related to the absorption (loss or gain) of the material.
The formula of $\varepsilon_{2}(\omega)$ is as follows \cite{ram},
\begin{align}
	\varepsilon_{2}(\omega) = &\frac{V e^2}{2\pi \hbar m^{2} \omega^{2}} \int d^{3}k \sum|\langle\psi_{c}|p|\psi_{v}\rangle|^{2} \times \nonumber
 \\
 &
 \delta (E_{c}-E_{v}-\hbar\omega),
\end{align}
where $\psi_{c}$ and $\psi_{v}$ are the wavefunctions for the conduction band and valence band respectively. $\hbar$ is the planck's constant and $\omega$ is the frequency of the photon. $p$ is the momentum operator.
The real part, $\varepsilon_{1}(\omega)$ of the dielectric constant can be obtained using the Kramers Kronig relation,
\begin{equation}
\varepsilon_{1}(\omega) = 1+\frac{2}{\pi} P \int_{0}^{\infty} \frac{\varepsilon_{2}(\omega')\omega'}{{\omega'}^2- \omega^2}d\omega',	
\end{equation}
where $P$ denotes the principle value of the integral.

The energy loss spectrum, $L(\omega)$ can be calculated using the following expression,
\begin{equation}
	L(\omega) = \frac{\varepsilon_{2}(\omega)}{\varepsilon_{1}^{2}(\omega)+\varepsilon_{2}^{2}(\omega)}.
\end{equation}
\begin{figure*}[!ht!]
	\begin{center}
		\phantomlabelabovecaption{(a)}{fig:7a}
		\phantomlabelabovecaption{(b)}{fig:7b}
		\phantomlabelabovecaption{(c)}{fig:7c}	
		\subfloat{\includegraphics[width=0.9\textwidth]{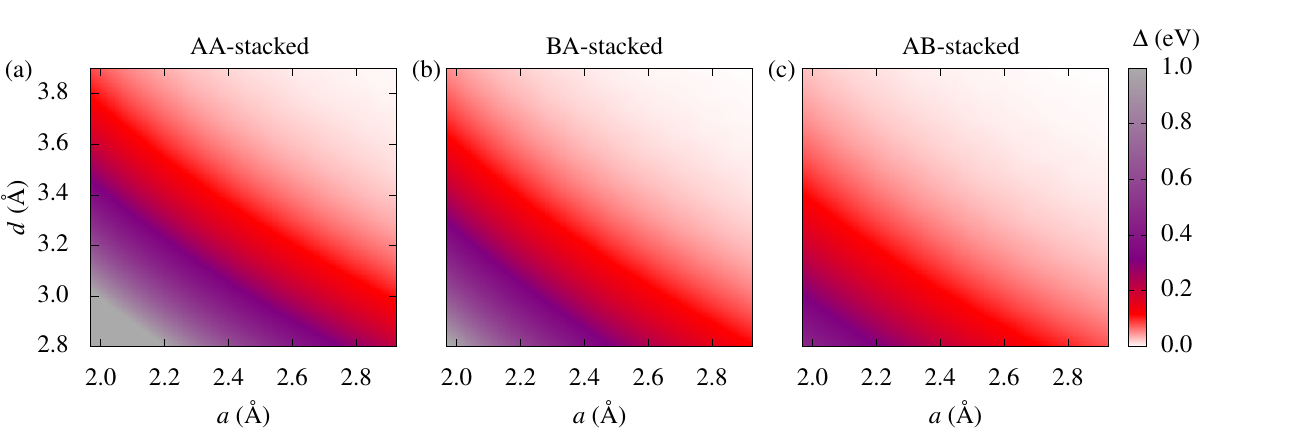}}
		\caption{(Color online) Energy gap variation, $\Delta$ (in units of eV) is plotted as a function of interlayer distance, $d$ (in units of \AA) and lattice constant, $a$ (in units of \AA) for different stacking geometries (a) AA-stacking, (b) BA-stacking, and (c) AB-stacking of graphene/hBN heterostructure. The color bar indicates the magnitude of the energy gap, $\Delta$ (in units of eV).}
		\label{fig:7}
	\end{center}
\end{figure*}
\begin{figure}[!t!]
	\begin{center}
		\subfloat{\includegraphics[trim=0 0 0 0,clip,width=0.4\textwidth]{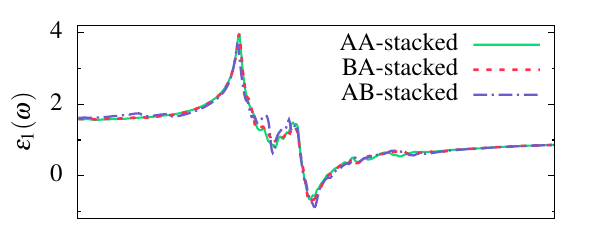}\label{fig:8a}}	\\ \vspace*{-1.04 cm}
		\subfloat{\includegraphics[trim=0 0 0 0,clip,width=0.4\textwidth]{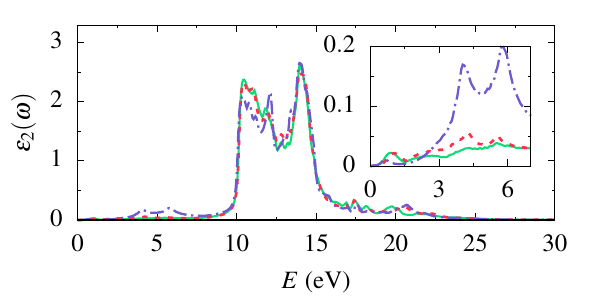}\label{fig:8b}}
		\caption{(Color online) Real part, $\varepsilon_{1}(\omega)$ and imaginary part, $\varepsilon_{2}(\omega)$ of dielectric constant are plotted as a function of photon energy, $E$ (in units of eV) for all three stackings of graphene/hBN heterostructure.}
		\label{fig:8}
	\end{center}
\end{figure}
\section{Results}\label{III}
In the following sections, we mainly discuss the electronic and optical properties of graphene/hBN heterostructure. We initially study the electronic band structure and the orbital contribution in the density of states (also known as the projected density of states) without any effects of strain. Then, we show how the energy gap induced by the hBN substrate gets modified under the influence of strain. Next, we investigate the optical properties and discuss the impact of strain on these properties.  
\subsection{Band structure and PDOS}\label{IIIA}
We initially calculate the minimum energy distances for three different stacking geometries considered in our model, as illustrated in Fig.~\ref{fig:1}. The total energy curve, $E_{\text{tot}}$ as a function of the interlayer distance between the graphene and hBN, $d$ is shown for AA-stacking (denoted by the green curve), BA-stacking (denoted by the red curve) and AB-stacking (denoted by the blue curve) in Fig.~\ref{fig:2}. We show that the minimum energy distances vary from one stacking to another, and it is found to be 3.51~\AA, 3.42~\AA, and 3.21~\AA~for AA-stacking, BA-stacking, and AB-stacking, respectively. This study also confirms that the total energy is the lowest for AB-stacking and is hence considered to be the most energetically favorable configuration. 
 
In Fig.~\ref{fig:3} and Fig.~\ref{fig:4}, we have shown the electronic band structure as well as the projected and the total density of states (DOS) of graphene/hBN heterostructure for all stacking geometries at their respective equilibrium distances. Due to the broken symmetry, a small direct band gap in the energy spectrum appears at the Dirac point, $K$ as depicted in Fig.~\ref{fig:3}. However, the band structure exhibits almost similar features (preserving the Dirac cone nature) for all the stacking geometries except the energy gap value at the $K$-point. At the equilibrium distance, the energy band gap measures 61 meV and 49 meV for AA-stacking and BA-stacking, respectively, whereas for AB-stacking, it corresponds to 60 meV. Hence, the largest band gap is found for the AA-stacking which arises due to the symmetry breaking of the two different sublattices (chiral) of the carbon atoms. Since the gap is very small, we have shown the zoomed view near the $K$-point in the inset of Fig~\ref{fig:3}. The DOS spectrum in Fig.~\ref{fig:4} shows that the major contribution arises from the $p$-orbital state of the C atom, whereas, for B and N atoms, it has zero contribution near the Fermi level. This is true for all three different configurations. Also, it can be clearly seen that the contribution from the $p$-orbital state of the C atom and the total DOS is almost equal in the vicinity of the Fermi level. 
\begin{figure*}[!ht!]
	\begin{center}
		\phantomlabelabovecaption{(a)}{fig:9a}
		\phantomlabelabovecaption{(b)}{fig:9b}
		\phantomlabelabovecaption{(c)}{fig:9c}
		\phantomlabelabovecaption{(d)}{fig:9d}
		\phantomlabelabovecaption{(e)}{fig:9e}
		\phantomlabelabovecaption{(f)}{fig:9f}		
		\subfloat{\includegraphics[width=0.8\textwidth]{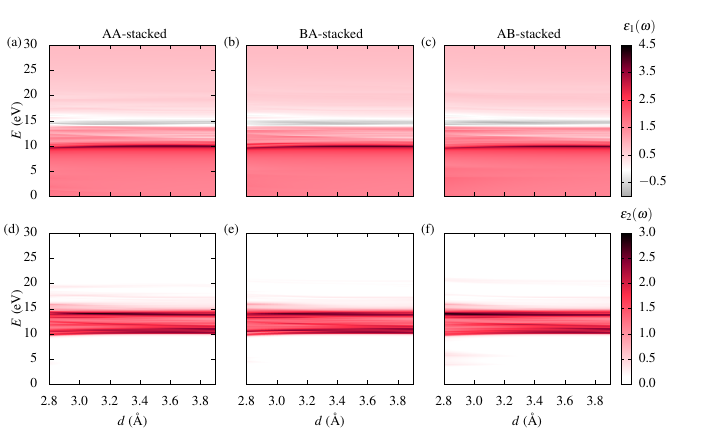}}	
		\caption{(Color online) Real part, $\varepsilon_{1}(\omega)$ and imaginary part, $\varepsilon_{2}(\omega)$ of dielectric constant are plotted as a function of interlayer distance, $d$ (in units of \AA) and photon energy, $E$ (in units of eV) for (a) and (d) AA-stacking, (b) and (e) BA-stacking, (c) and (f) AB-stacking of graphene/hBN heterostructure. The color bars in the $z$-direction represent the value of dielectric constants in both real ($\varepsilon_{1}(\omega)$) and imaginary ($\varepsilon_{2}(\omega)$) parts respectively.}
		\label{fig:9}
	\end{center}
\end{figure*}
\begin{figure*}[!ht!]
	\begin{center}
		\phantomlabelabovecaption{(a)}{fig:10a}
		\phantomlabelabovecaption{(b)}{fig:10b}
		\phantomlabelabovecaption{(c)}{fig:10c}	
		\subfloat{\includegraphics[width=0.75\textwidth]{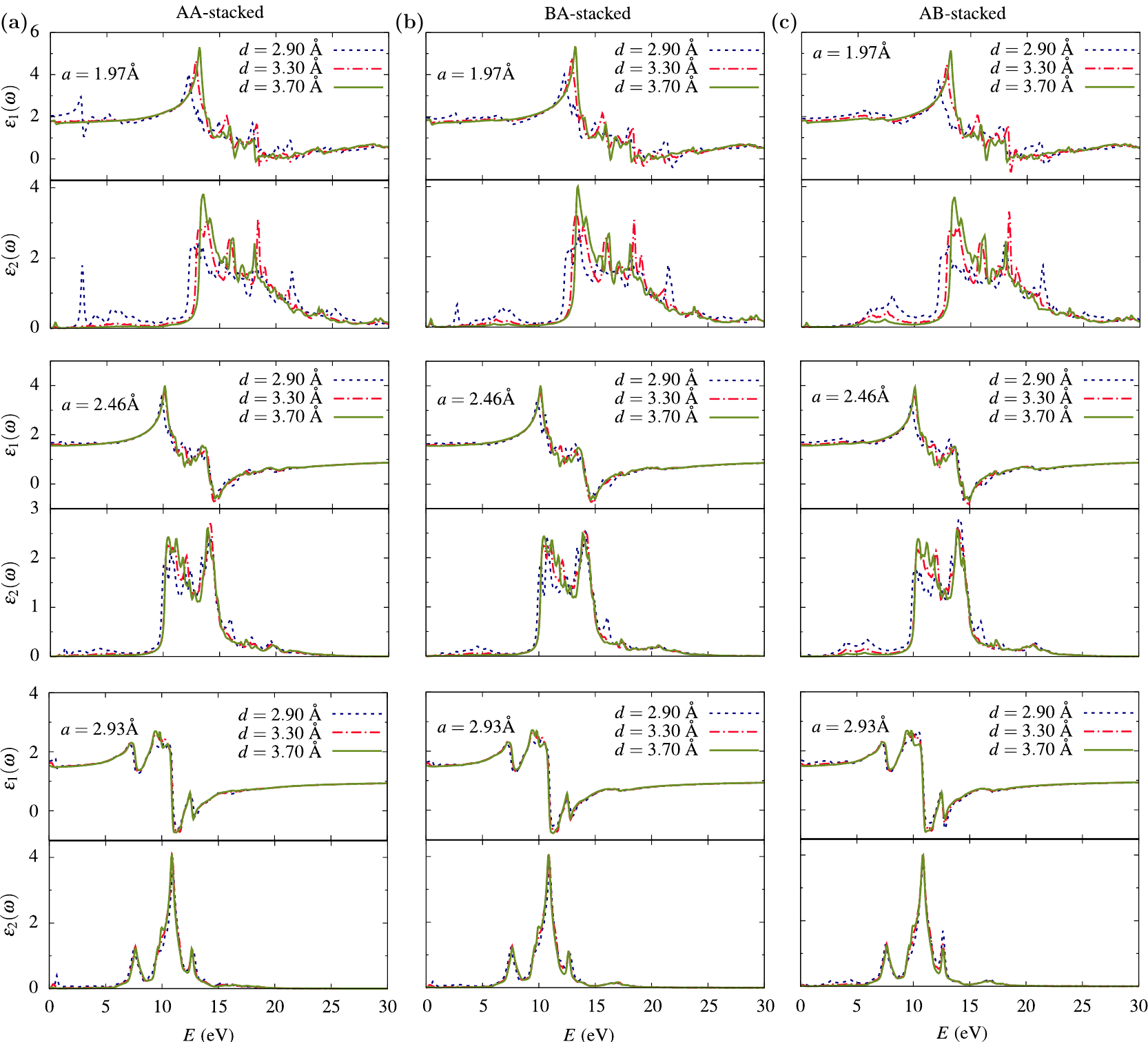}}
		\caption{(Color online) Real part, $\varepsilon_{1}(\omega)$ and imaginary part, $\varepsilon_{2}(\omega)$ of dielectric constant are plotted as a function of photon energy, $E$ (in units of eV) for different interlayer distance, $d$ (in units of \AA) and lattice constant, $a$ (in units of \AA) for (a) AA-stacking (left panel), (b) BA-stacking (middle panel), and (c) AB-stacking (right panel) of graphene/hBN heterostructure.}
		\label{fig:10}
	\end{center}
\end{figure*}
\begin{figure}[!ht!]
\begin{center}
  \subfloat{\includegraphics[width=0.35\textwidth]{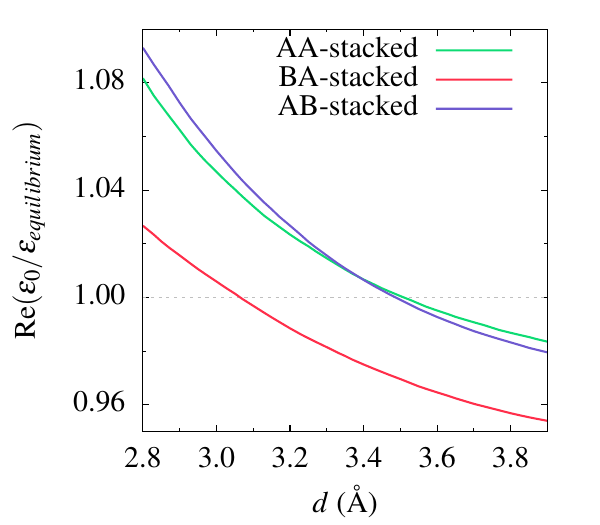}}
\caption{(Color online) Real part of static dielectric constant is shown as a function of interlayer distance, $d$ (in units of \AA) for all three stackings of graphene/hBN heterostructure.}
\label{fig:11}
\end{center}
\end{figure}
\begin{figure*}[!ht!]
	\begin{center}
	\phantomlabelabovecaption{(a)}{fig:AA_4}
	\phantomlabelabovecaption{(b)}{fig:BA_4}
	\phantomlabelabovecaption{(c)}{fig:AB_4}		
	\subfloat{\includegraphics[width=0.33\textwidth]{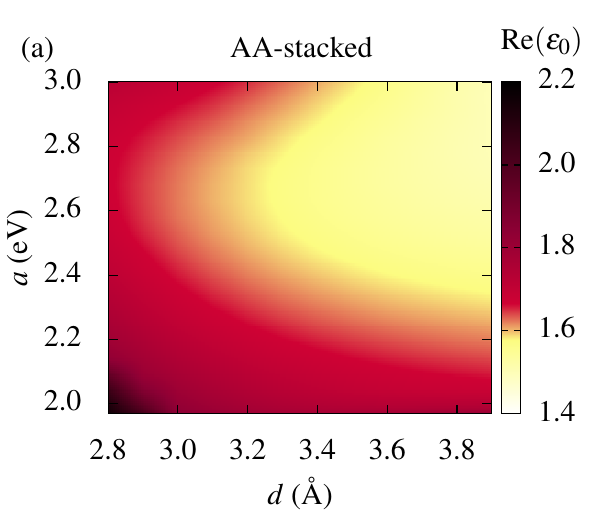}\label{fig:12a}}	
	\subfloat{\includegraphics[width=0.33\textwidth]{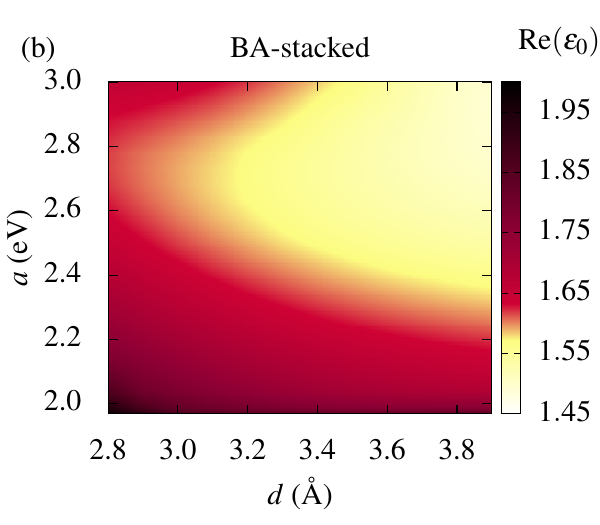}\label{fig:12b}}
	\subfloat{\includegraphics[width=0.33\textwidth]{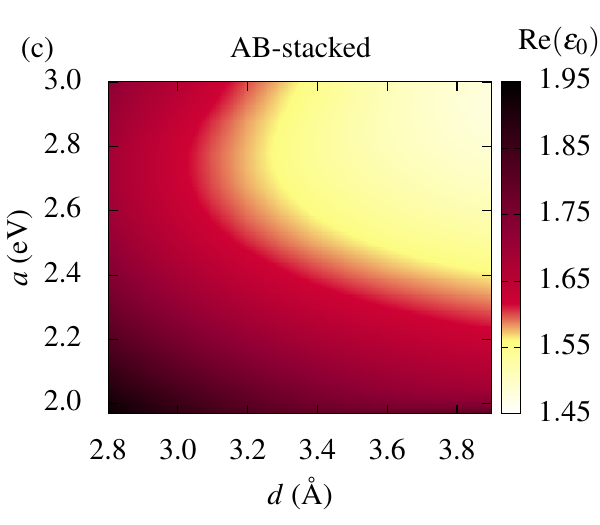}\label{fig:12c}}
		\caption{(Color online) Real part of static dielectric constant, Re($\varepsilon_0$) is shown as a function of interlayer distance, $d$ (in units of \AA) and lattice constant, $a$ (in units of \AA) for (a) AA-stacking, (b) BA-stacking, and (c) AB-stacking of graphene/hBN heterostructure.}
		\label{fig:12}
	\end{center}
\end{figure*}
\subsection{Strain induced band gap}\label{IIIB} 
To validate our model, we demonstrate the variation in the energy gap, $\Delta_\text{gap}$ as a function of interlayer distance, $d$ for all the stacking geometries which has been reported previously \cite{Giovannetti}. This is depicted in Fig.~\ref{fig:5a}. We observe that the tunability of the energy band gap depends on $d$ and it decreases with the increasing distance between the graphene and hBN. However, $\Delta_\text{gap}$ also varies as a function of $d$ for all the individual stacking geometries. As the distance increases, the sublattice symmetry breaking reduces which eventually leads to the closing of the band gap for all the stacking geometries.

Next, we demonstrate the variation of the energy band gap in the presence of uniaxial strain (applied in one direction), where we choose the lattice constant of the system as the tuning parameter. Fig.~\ref{fig:5b} shows the dependence of the energy band gap, $\Delta_\text{gap}$ on the lattice constant, $a$ for all three stacking geometries. The green and the red curves (corresponding to AA and BA stacking, respectively) both decrease monotonically as $a$ increases. Conversely, the blue curve (corresponding to AB stacking) initially shows a significant gap at very small values of $a$ and subsequently exhibits a more rapid decrease as compared to the other two curves with the increasing $a$ (see Fig.~\ref{fig:5b}). Therefore, the tuning of the band gap in the system can be effectively obtained in the presence of uniaxial strain, resulting in a transition from a semiconductor state to a semimetal state. 

Next, we have shown the band structure and the DOS by tuning both $a$ and $d$ for AA-stacking (left panel), BA-stacking (middle panel), and AB-stacking (right panel) in Figs.~\ref{fig:6a}-\ref{fig:6c} respectively. When $a=1.97$ \AA~and $d=2.90$ \AA~(both are very small), the energy band gap becomes very large for an AA-stacking as compared to BA and AB stacking and it decreases with the increasing $d$ value (see the top panel in Fig.~\ref{fig:6}). As we increase the $a$ value (say $a=2.46$ \AA), the energy gap starts decreasing in a previous manner for all three stackings and eventually tends to close when both $a$ and $d$ are large (say, $a=2.93$ \AA~and $d=3.70$ \AA)~as can be seen from the bottom panel of Fig.~\ref{fig:6}. 

In Fig.~\ref{fig:7a}-\ref{fig:7c}, we show the color map of the calculated band gap variation, $\Delta$ as a function of both the parameters $d$ and $a$ for the AA-stacking, BA-stacking, and AB-stacking respectively. The color bar indicates the magnitude of the energy gap value, which varies quantitatively from one stacking to another. However, the gap, $\Delta$ shows the maximum value for AA-stacking when both $d$ and $a$ are very small, as compared to the other two stackings. As we further increase the value of both $d$ and $a$, the band gap reduces and tends to almost zero value for all the stacking geometries as can be seen from Fig.~\ref{fig:7}.    
\subsection{Optical properties}\label{IIIC}
In this section, we investigate the optical response of pristine graphene/hBN heterostructure for a wide range of photon energy spectrum. Since the electronic structure of graphene is significantly influenced when combined with hBN resulting in a band gap, it becomes intriguing to investigate the impact of the graphene/hBN heterostructure on optical properties. Several studies of the total optical response from graphene/hBN heterostructures have been reported for a wide spectral range \cite{jia,tok,gx,rig}. In Fig.~\ref{fig:8}, we have shown the real ($\varepsilon_{1}(\omega)$) and the imaginary ($\varepsilon_{2}(\omega)$) parts (also known as relative permittivity) for the z-components of the dielectric constant as a function of photon energy, $E$ for AA-stacking, BA-stacking, and AB-stacking at their equilibrium distances. The behavior of $\varepsilon_{1}(\omega)$ (top panel) is qualitatively the same for all three different stackings. The value of the $\varepsilon_{1}(\omega)$ at very low frequency (which corresponds to zero photon energy) is referred to as static dielectric constant, $\varepsilon_{0}$. However, the value of $\varepsilon_{0}$ is found to be maximum for AB-stacking ($\varepsilon_{0}$ = 1.61), whereas for AA and BA stackings, $\varepsilon_{0}$ takes almost the same value, i.e., 1.57 and 1.58 respectively. The maximum value of $\varepsilon_{0}$ indicates that AB stacking has higher electromagnetic storage capacity than AA and BA stacking. 

The imaginary part of the dielectric constant, $\varepsilon_{2}(\omega)$ is directly associated with the energy absorption of the system. For a direct band gap semiconductor, the vertical transition from the valence band maximum to the conduction band minimum gives the first absorption peak in the spectrum. For all the stackings, the first peak in $\varepsilon_{2}(\omega)$ (bottom panel) is observed in the infrared regime which can be clearly seen from the inset of Fig.~\ref{fig:8}. The peaks in $\varepsilon_{2}(\omega)$ increases in the ultra-violet region ($\approx$ 3-6 eV) for AB stacking (shown by the blue curve) as compared to the other two stackings. However, the absorption spectra show more pronounced and broadened peaks within the energy range $10-15$ eV (which corresponds to 80-130 nm wavelength range) in the ultra-violet region. This result implies that the graphene on hBN absorbs light over a broad frequency range in this region, whereas the absorption is relatively low at infrared frequencies. 
\begin{figure}[!ht!]
	\begin{center}
		\subfloat{\includegraphics[width=0.33\textwidth]{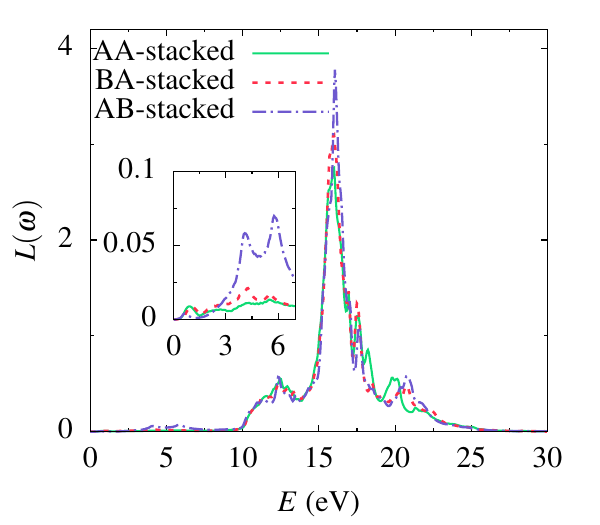}}
		\caption{(Color online) Dielectric loss spectrum, $L(\omega)$ is plotted as a function of photon energy, $E$ (in units of eV) for all three stackings of graphene/hBN heterostructure. The inset shows the zoomed view in the low-energy region.}
		\label{fig:13}
	\end{center}
\end{figure}
\begin{figure*}[!ht!]
	\begin{center}	
         \phantomlabelabovecaption{(a)}{fig:14a}
		\phantomlabelabovecaption{(b)}{fig:14b}
		\phantomlabelabovecaption{(c)}{fig:14c}	
        \subfloat{\includegraphics[width=0.8\textwidth]{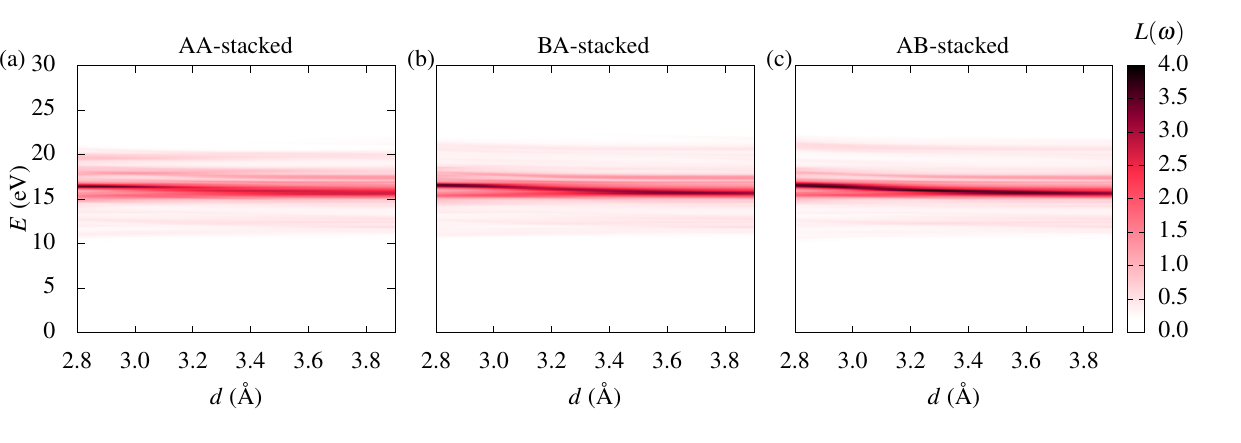}}	
	\caption{(Color online) Dielectric loss spectrum, $L(\omega)$ is shown as a function of interlayer distance, $d$ (in units of \AA) and photon energy, $E$ (in units of eV) for (a) AA-stacking, (b) BA-stacking, and (c) AB-stacking of graphene/hBN heterostructure. The color bar in the $z$-direction represents the value of the loss function.}
		\label{fig:14}
	\end{center}
\end{figure*}
\begin{figure*}[!ht!]
	\begin{center}
		\phantomlabelabovecaption{(a)}{fig:15a}
		\phantomlabelabovecaption{(b)}{fig:15b}
		\phantomlabelabovecaption{(c)}{fig:15c}	
		\subfloat{\includegraphics[width=0.8\textwidth]{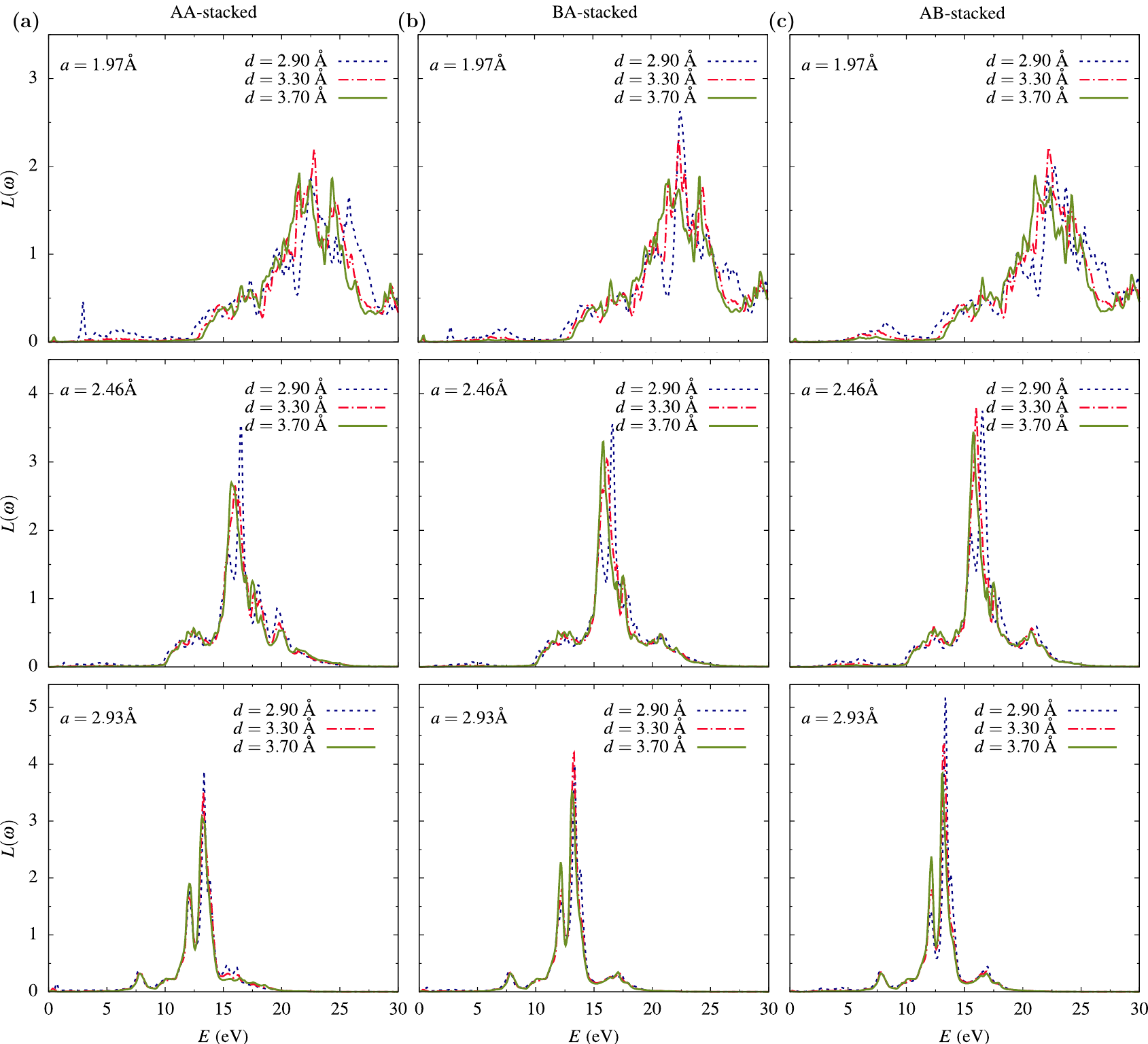}}
		\caption{(Color online) Dielectric loss spectrum, $L (\omega)$ is plotted as a function of photon energy, $E$ (in units of eV) for different interlayer distance, $d$ (in units of \AA) and lattice constant, $a$ (in units of \AA) for (a) AA-stacking (left panel), (b) BA-stacking (middle panel), and (c) AB-stacking (right panel) of graphene/hBN heterostructure.}
		\label{fig:15}
	\end{center}
\end{figure*} 
\begin{figure*}[!ht!]
	\begin{center}
		\phantomlabelabovecaption{(a)}{fig:16a}
		\phantomlabelabovecaption{(b)}{fig:16b}
		\phantomlabelabovecaption{(c)}{fig:16c}
		\subfloat{\includegraphics[width=0.9\textwidth]{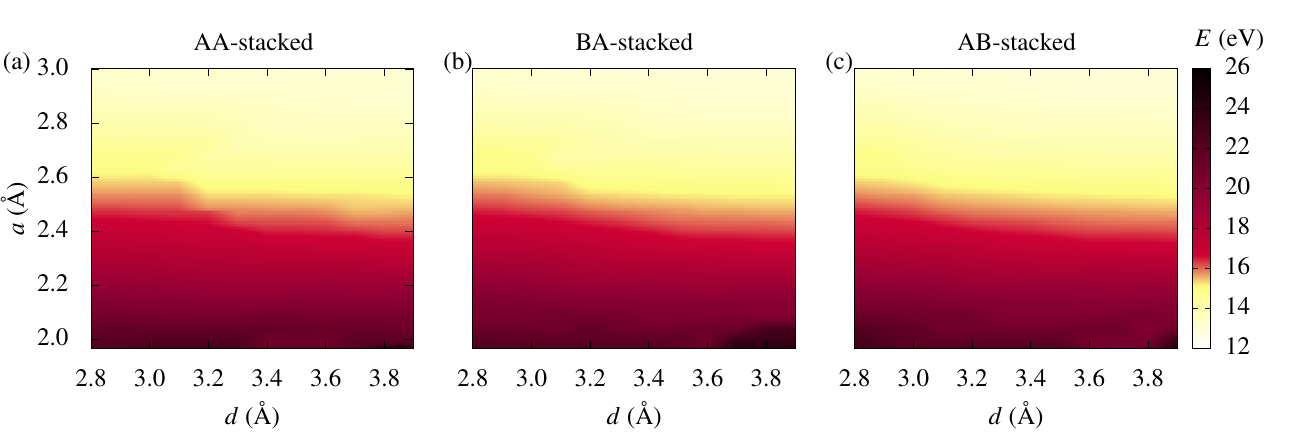}}
		\caption{(Color online) Maximum electron energy loss is shown as a function of interlayer distance, $d$ (in units of \AA) and lattice constant, $a$ (in units of \AA) for (a) AA-stacking, (b) BA-stacking, and (c) AB-stacking of graphene/hBN heterostructure.}
		\label{fig:16}
	\end{center}
\end{figure*}
\begin{figure}[h]
\begin{center}
  \subfloat{\includegraphics[width=0.33\textwidth]{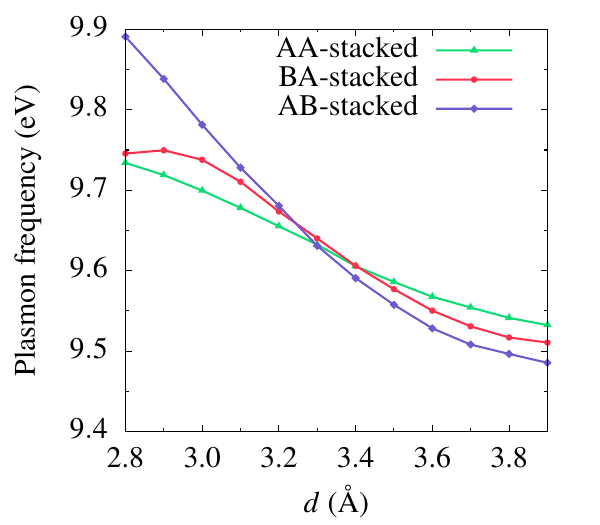}}
\caption{(Color online) Plasmon frequency (in units of eV) is shown as a function of interlayer distance, $d$ (in units of \AA) for all three stackings of graphene/hBN heterostructure.}
\label{fig:17}
\end{center}
\end{figure}
\subsubsection{Effects of strain}\label{d}
In this subsection, we introduce the application of strain to study the optical properties of the graphene/hBN heterostructure considering all the stacking geometries. In Fig.~\ref{fig:9}, we present the color map of the calculated real part, $\varepsilon_{1}(\omega)$ and imaginary part, $\varepsilon_{2}(\omega)$ ($z$-components) of the dielectric constant as a function of photon energy, $E$ and the interlayer distance, $d$ for AA-stacking (Figs.~\ref{fig:9a} and \ref{fig:9d}), BA-stacking (Figs.~\ref{fig:9b} and \ref{fig:9e}), and AB-stacking (Figs.~\ref{fig:9c} and \ref{fig:9f}). The real part, $\varepsilon_{1}(\omega)$ does not show strong variation as a function of both $E$ and $d$, which is true for all stacking geometries. The maximum peak value of $\varepsilon_{1}(\omega)$ is observed to occur at an energy of approximately 10 eV for all three stackings. For the imaginary part, the value of $\varepsilon_{2}(\omega)$ shows the maximum absorption peaks between the energy range $10-15$ eV (which corresponds to wavelength $\sim 80-124$ nm) in the ultra-violet region. However, the qualitative nature of the imaginary part of the dielectric constant ($\varepsilon_{2}(\omega)$) is similar for all the stacking geometries. 

Further, we have shown the variation of the real ($\varepsilon_{1}(\omega)$) and imaginary ($\varepsilon_{2}(\omega)$) parts of the dielectric constant with the interlayer distance, $d$ and the lattice constant, $a$ for AA-stacking (left panel), BA-stacking (middle panel), and AB-stacking (right panel) in Fig.~\ref{fig:10}. For an AA-stacked system (see Fig.~\ref{fig:10a}), $\varepsilon_{1}(\omega)$ at very low energy decreases as we increase the interlayer distance, $d$ for a fixed $a$ value. However, when we increase the $a$ value (say, $a=2.93$\AA), $\varepsilon_{1}(\omega)$ decreases more with the increasing values of $d$. The imaginary part, $\varepsilon_{2}(\omega)$ shows the maximum peak in the low energy region for AA-stacking with a small $a$ and large $d$ value (say, $a=1.97$\AA~and $d=3.70$\AA) (see top of Fig.~\ref{fig:10a}). When we increase the $a$ value (say, $a=2.93$\AA), the maximum peak is observed for a small $d$ value, and the peak decreases with the increasing $d$ (see bottom of Fig.~\ref{fig:10a}). A similar qualitative feature is observed for the other two stacking geometries, that is, BA-stacking and AB-stacking (as seen from Figs.~\ref{fig:10b} and \ref{fig:10c}). However, with the increasing $d$ value, the maximum absorption peak obsevered across a broad range shifted towards the lower energy for all three stackings. 

Next, we have shown how the real part of static dielectric constant, $\varepsilon_{0}$ varies as a function of interlayer distance $d$ in Fig.~\ref{fig:11}. It is evident that the value of the Re($\varepsilon_{0}$) decreases gradually with the increasing distance, $d$. Previous studies have indicated that interlayer coupling is enhanced under pressure, leading to an overall redshift in absorption and an increase in the dielectric constant \cite{jung}. However, the maximum value of Re($\varepsilon_{0}$) is observed for AB-stacking when compared to the other two stackings at small $d$ values. Furthermore, the value of Re($\varepsilon_{0}$) for AA and AB stacking is more comparable than that of BA-stacking. In Fig.~\ref{fig:12a}-\ref{fig:12c}, we have shown the color map for the real part of the static dielectric constant, Re($\varepsilon_{0}$) where we have tuned both the parameter $a$ and $d$ simultaneously for AA-stacking, BA-stacking and AB-stacking geometries respectively. When both $a$ and $d$ are small, the value of Re($\varepsilon_{0}$) becomes maximum for all stacking geometries. Further, it shows the same value for either of the cases (i) $d$ increases with a small $a$ value or (ii) $a$ increases with a small $d$ value. However, when both $d$ and $a$ increase, the value of Re($\varepsilon_{0}$) significantly decreases.
\subsubsection{Loss spectrum}\label{e}
In this subsection, we compute the dielectric loss spectra, $L(\omega)$ representing the energy loss of a fast-moving electron within the system, as a function of photon energy, $E$, for all three different stacking without any strain, as depicted in Fig.~\ref{fig:13}. The $L(\omega)$ spectra exhibit small intensity peaks in the infrared region, with a subsequent increase in intensity in the ultraviolet region. Notably, in the ultraviolet region, the peak intensity for the AB-stacked system is greater than that of the other two stackings. Though the intensity of the peak varies from one stacking to another, the maximum value of the peak occurs at the same photon energy ($\sim 16$ eV) for all three stackings as seen from Fig.~\ref{fig:13}. In Figs.~(\ref{fig:14a}-\ref{fig:14c}), we have shown the electron energy loss spectrum as a function of $d$ and photon energy, $E$ for AA-stacking, BA-stacking, and AB-stacking respectively. When $d$ is small, the maximum electron energy loss is high, and it decreases with the increasing value of $d$ and becomes almost constant for large $d$. 

Next, we have plotted the electron energy loss spectrum, $L(\omega)$ as a function of photon energy, $E$ with the variation of $d$ and $a$ for AA-stacking (left panel), BA-stacking (middle panel) and AB-stacking (right panel) as shown in Figs.~(\ref{fig:15a}-\ref{fig:15c}) respectively. When $a=1.97$ \AA, the maximum energy loss is observed for a wide range of spectrum in the ultra-violet region for all stacking geometries. As we increase the $a$ value (say, $2.46$ \AA), the maximum energy loss peak in $L(\omega)$ spectra shifts towards the lower energy for all three stackings. Moreover, it can be seen that the maximum loss occurs at different photon energies for different $d$ values, and it actually shifts towards the lower energy with the increasing $d$. However, when $a$ becomes 2.93 \AA, the maximum loss peak in $L(\omega)$ shifts more towards the lower energy and remains almost the same for all $d$ values. In Figs.~(\ref{fig:16a}-\ref{fig:16c}), we have shown the maximum energy loss of the electron as a function of both $a$ and $d$ for AA-stacking, BA-stacking, and AB-stacking respectively. The maximum energy loss is high for BA stacking when $d$ is large and $a$ is small (see Fig.~\ref{fig:16b}). However, the loss decreases with the increasing $a$ value irrespective of the change in interlayer distance which is true for all the stacking geometries.

This peaks in $L(\omega)$ represent the characteristic behaviors associated with the plasma oscillations, and the corresponding frequencies are the so-called plasma frequencies, which have been described in the following Sec.~\ref{f}.   
\subsubsection{surface plasmon}\label{f}
Although graphene is an excellent plasmonic material due to its low loss and high tunability, it is often combined with other 2D materials \cite{mak, merano} to further enhance its properties. Typically, the wavelength of graphene plasmon is two orders of magnitude smaller than the wavelength of light. However,  the combination of graphene and hBN can modify the plasmonic behavior of graphene and enable the engineering of optical resonances. For example, when graphene is combined with hBN, the wavelength of plasmon changes and approaches the same order of magnitude as the wavelength of light. Optically, they are expected to produce unusual plasmonic behavior in the case of an aligned lattice \cite{toma}. In Fig.~\ref{fig:17}, the dependence of the plasmon frequency of the graphene/hBN system on the interlayer distances is illustrated for all three different stackings. In fact, the plasmon frequency falls in the ultra-violet region. It can be seen that the plasmon frequency decreases as a function of interlayer distance $d$ which is true for all the stacking geometries. However, depending on the stacking configurations, the frequency curve may either increase or decrease. Notably, the frequency decreases more rapidly with increasing interlayer distances for AB stacking (denoted by the blue curve) compared to AA and BA stackings. Also, there is a crossover in the spectrum where all three of them meet at the same interlayer distance $d$. Nevertheless, it is possible to tune the plasmon frequency of the system with the interlayer distance, which may allow control over the graphene surface plasmon with incident light. 
\section{Summary}\label{IV}
In conclusion, we have studied the tunability of the electronic band gap and the optical properties of the graphene/hBN heterostructure under the application of strain from the first principles calculations. We observed that the band gap decreases as both the interlayer distance and lattice constant increase, eventually approaches to zero value for large $d$ and $a$ value. We demonstrated that significant tunability of the band gap can be achieved by simultaneously adjusting both the interlayer distance and lattice constant. Further, we demonstrated that the optical properties show unique behavior owing to the direct band gap within the system. We computed real and imaginary parts of the dielectric constant and the loss spectrum for graphene/hBN heterostructure. Our findings revealed that the value of the dielectric constant of the graphene/hBN heterostruct becomes maximum when both $d$ and $a$ are small which is true all stacking geometries. Moreover, we found that graphene/hBN plasmon offers superior tunability, which makes it suitable for optoelectronic devices. Hence, this study underscores the significant interplay between strain and the optical properties of graphene on hBN heterostructures.
\begin{acknowledgments}
The authors acknowledge the support provided by the KEPLER computing facility maintained by the Department of Physical Sciences, IISER Kolkata. P.~S. thanks Dr. T. Samui for the helpful discussions.
\end{acknowledgments}


\begin{thebibliography}{99}
\bibitem{geim2} A. K. Geim and I. V. Grigorieva, Van der Waals heterostructures, \href{https://doi.org/10.1038/nature12385}{Nature {\bf 499}, 419 (2013)}.

\bibitem{la} L. Ponomarenko, A. Geim, A. Zhukov \textit{et al.}, Tunable metal–insulator transition in double-layer graphene heterostructures, \href{https://doi.org/10.1038/nphys2114}{Nature Phys. {\bf 7}, 958 (2011)}.

\bibitem{brit} L. Britnell, R. V. Gorbachev, and R. Jalil \textit{et al.}, Field-Effect Tunneling Transistor Based on Vertical Graphene Heterostructures, \href{https://doi.org/10.1126/science.1218461}{Science {\bf 335}, 947 (2012)}.

\bibitem{haigh} S. J. Haigh, A. Gholinia, R. Jalil \textit{et al.}, Cross-sectional imaging of individual layers and buried interfaces of graphene-based heterostructures and superlattices, \href{https://doi.org/10.1038/nmat3386}{Nature Mater. {\bf 11}, 764 (2012)}.

\bibitem{Giovannetti} G. Giovannetti, P. A. Khomyakov, G. Brocks, P. J. Kelly, and J. van den Brink, Substrate-induced band gap in graphene on hexagonal boron nitride: Ab initio density functional calculations, \href{https://doi.org/10.1103/PhysRevB.76.073103}{Phys. Rev. B {\bf 76}, 073103 (2007)}.

\bibitem{pakdel} A. Pakdel, C. Zhi, Y. Bando, and D. Golberg, Low-dimensional boron nitride nanomaterials, \href{https://doi.org/10.1016/S1369-7021(12)70116-5}{Mater. Today {\bf 15}, 256 (2012)}.

\bibitem{tani} K. Watanabe, T. Taniguchi, and H. Kanda, Direct-bandgap properties and evidence for ultraviolet lasing of hexagonal boron nitride single crystal, \href{https://doi.org/10.1038/nmat1134}{Nat. Mater. {\bf 3}, 404 (2004)}.

\bibitem{pp} A. Kumar, P. K. Panigrahi, Origin of Lattice Spin in Graphitic Systems,  \href{https://doi.org/10.48550/arXiv.1512.06037}{arXiv:1512.06037}.

\bibitem{wa} X. Wang, Y. Ouyang, X. Li, H. Wang, J. Guo, and H. Dai, Room-Temperature All-Semiconducting Sub-10-nm Graphene Nanoribbon Field-Effect Transistors, \href{https://doi.org/10.1103/PhysRevLett.100.206803}{Phys. Rev. Lett. {\bf 100}, 206803 (2008)}.

\bibitem{balog} R. Balog, B. J\o rgensen, and L. Nilsson \textit{et al.}, Bandgap opening in graphene induced by patterned hydrogen adsorption, \href{https://doi.org/10.1038/nmat2710}{Nat. Mater. {\bf 9}, 315 (2010)}.

\bibitem{kang} Y.-J. Kang, J. Kang, and K. J. Chang, Electronic structure of graphene and doping effect on SiO$_{2}$, \href{https://doi.org/10.1103/PhysRevB.78.115404}{Phys. Rev. B {\bf 78}, 115404 (2008)}.

\bibitem{maas} T. Maassen, F. K. Dejene, M. H. D. Guimar\~aes, C. J\'ozsa, and B. J. van Wees, Comparison between charge and spin transport in few-layer graphene, \href{https://doi.org/10.1103/PhysRevB.83.115410}{Phys. Rev. B {\bf 83}, 115410 (2011)}.

\bibitem{fan} X. F. Fan, W. T. Zheng, V. Chihaia, Z. X. Shen, and J.-L. Kuo, Interaction between graphene and the surface of SiO$_{2}$, \href{https://doi.org/10.1088/0953-8984/24/30/305004}{J. Phys. Condens. Matter {\bf 24}, 305004 (2012)}.

\bibitem {dean} C. R. Dean, A. F. Young, I. Meric, C. Lee, L. Wang, S. Sorgenfrei, K. Watanabe, T. Taniguchi, P. Kim, K. L. Shepard, J. Hone, Boron nitride substrates for high-quality graphene electronics, \href{https://doi.org/10.1038/nnano.2010.172}{Nat. Nanotech. {\bf 5}, 722 (2010)}.

\bibitem{son} S.-K. Son, M. \v{S}i\v{s}kins, and C. Mullan \textit{et al.}, Graphene hot-electron light bulb: incandescence from hBN-encapsulated graphene in air, \href{https://doi.org/10.1088/2053-1583/aa97b5}{2D Mater. {\bf 5}, 011006 (2017)}.

\bibitem{band} D. A. Bandurin, D. Svintsov, I. Gayduchenko \textit{et al.}, Resonant terahertz detection using graphene plasmons, \href{https://doi.org/10.1038/s41467-018-07848-w}{Nat. Commun. {\bf 9}, 5392 (2018)}.

\bibitem{viti} L. Viti, D. G. Purdie, A. Lombardo, A. C. Ferrari, and M. S. Vitiello, HBN-Encapsulated, Graphene-based, Room-temperature Terahertz Receivers, with High Speed and Low Noise, \href{https://doi.org/10.1021/acs.nanolett.9b05207}{Nano Lett. {\bf 20}, 3169 (2020)}.

\bibitem{song} S.-B. Song, S. Yoon, S. Y. Kim \textit{et al.}, Deep-ultraviolet electroluminescence and photocurrent generation in graphene/hBN/graphene heterostructures, \href{https://doi.org/10.1038/s41467-021-27524-w}{Nat. Commun. {\bf 12}, 7134 (2021)}.

\bibitem{with} F. Withers, O. D. Pozo-Zamudio, A. Mishchenko \textit{et al.}, Light-emitting diodes by band-structure engineering in van der Waals heterostructures,  \href{https://doi.org/10.1038/nmat4205}{Nat. Mater. {\bf 14}, 301 (2015)}.

\bibitem{shi} R. J. Shiue, Y. Gao, Y. Wang \textit{et al.}, High-Responsivity Graphene–Boron Nitride Photodetector and Autocorrelator in a Silicon Photonic Integrated Circuit, \href{https://doi.org/10.1021/acs.nanolett.5b02368}{Nano Lett {\bf 15}, 7288 (2015)}.

\bibitem{meng} J.-H. Meng, X. Liu , X.-W. Zhang \textit{et al.}, Interface engineering for highly efficient graphene-on-silicon Schottky junction solar cells by introducing a hexagonal boron nitride interlayer, \href{https://doi.org/10.1016/j.nanoen.2016.08.028}{Nano Energy {\bf 28}, 44 (2016)}.

\bibitem{choi} S.-M. Choi, S.-H. Jhi, and Y.-W. Son, Effects of strain on electronic properties of graphene, \href{https://doi.org/10.1103/PhysRevB.81.081407}{Phys. Rev. B  {\bf 81}, 081407(R) (2010)}.

\bibitem{j} J. Zhang, X. Lang, Y. Zhu, and Q. Jiang, Strain tuned InSe/MoS$_{2}$ bilayer van der Waals heterostructures for photovoltaics or photocatalysis, \href{https://doi.org/10.1039/C8CP02997K}{Phys. Chem. Chem. Phys. {\bf 20}, 17574–17582 (2018)}.

\bibitem{s} S. Postorino {\textit et al.}, Strain-induced effects on the electronic properties of 2D materials, \href{https://doi.org/10.1177/1847980420902569}{Nanomater. Nanotechnol. {\bf 10}, 1847980420902569 (2020)}.

\bibitem{nair} R. R. Nair, P. Blake, A. N. Grigorenko, K. S. Novoselov, T. J. Booth, T. Stauber, N. M. R. Peres, A. K. Geim, Fine Structure Constant Defines Visual Transparency of Graphene, \href{https://doi.org/10.1126/science.1156965}{Science {\bf 320} 1308 (2008)}.

\bibitem{tok} A. N. Toksumakov, G. A. Ermolaev, M. K. Tatmyshevskiy \textit{et al.}, Anomalous optical response of graphene on hexagonal boron nitride substrates,  \href{https://doi.org/10.1038/s42005-023-01129-9}{Commun. Phys. {\bf 6}, 13 (2023)}.

\bibitem{peres} T. Stauber, N. M. R. Peres, and A. K. Geim, Optical conductivity of graphene in the visible region of the spectrum, \href{https://doi.org/10.1103/PhysRevB.78.085432}{Phys. Rev. B {\bf 78}, 085432 (2008)}.

\bibitem{slot} G. J. Slotman, M. M. van Wijk, P.-L. Zhao, A. Fasolino, M. I. Katsnelson, and S. Yuan, Effect of Structural Relaxation on the Electronic Structure of Graphene on Hexagonal Boron Nitride, \href{https://doi.org/10.1103/PhysRevLett.115.186801}{Phys. Rev. Lett. {\bf 115}, 186801 (2015)}.

\bibitem{cate} A. Catellani, M. Posternak, A. Baldereschi, and A. J. Freeman, Bulk and surface electronic structure of hexagonal boron nitride, \href{https://doi.org/10.1103/PhysRevB.36.6105} {Phys. Rev. B {\bf 36}, 6105 (1987)}.

\bibitem{castro} A. H. Castro Neto, F. Guinea, N. M. R. Peres, K. S. Novoselov, and A. K. Geim, The electronic properties of graphene, \href{https://doi.org/10.1103/RevModPhys.81.109}{Rev. Mod. Phys. {\bf 81}, 109 (2009)}.

\bibitem{zoll} K. Zollner, M. Gmitra, and J. Fabian, Heterostructures of graphene and hBN: Electronic, spin-orbit, and spin relaxation properties from first principles, \href{https://doi.org/10.1103/PhysRevB.99.125151}{Phys. Rev. B {\bf 99}, 125151 (2019)}.

\bibitem{jose} P. San-Jose, A Guti\'errez-Rubio, M. Sturla and F. Guinea, Electronic structure of spontaneously strained graphene on hexagonal boron nitride
, \href{https://doi.org/10.1103/PhysRevB.90.115152}{Phys. Rev. B {\bf 90}, 115152 (2014)}.

\bibitem{balu} R. Balu, X. Zhong, R. Pandey, and S. P. Karna, Effect of electric field on the band structure of graphene/boron nitride and boron nitride/boron nitride bilayers, \href{https://doi.org/10.1063/1.3679174}{Appl. Phys. Lett. {\bf 100} 052104 (2012)}.

\bibitem{san} P. San-Jose, A Guti\'errez-Rubio, M. Sturla and F. Guinea, Spontaneous strains and gap in graphene on boron nitride
, \href{https://doi.org/10.1103/PhysRevB.90.075428}{Phys. Rev. B {\bf 90}, 075428 (2014)}. 

\bibitem{gia} P. Giannozzi, S. Baroni, N. Bonini \textit{et al.}, QUANTUM ESPRESSO: A modular and open-source software project for quantum simulations of materials, \href{https://doi.org/10.1088/0953-8984/21/39/395502}{J. Phys. Condens. Matter {\bf 21}, 395502 (2009)}.

\bibitem{ram} KR Babu, CB Lingam, S Auluck, SP Tewari, G Vaitheeswaran, Structural, thermodynamic and optical properties of MgF$_{2}$ studied from first-principles theory, \href{https://doi.org/10.1016/j.jssc.2010.11.025}{Journal of Solid State Chemistry {\bf 184}, 343-350 (2011)}.

\bibitem{jia} Y. Jia, H. Zhao, Q. Guo, X. Wang, H. Wang, and F. Xia, Tunable Plasmon–Phonon Polaritons in Layered Graphene–Hexagonal Boron Nitride Heterostructures, \href{https://doi.org/10.1021/acsphotonics.5b00099}{ACS Photonics {\bf 2}, 907-912 (2015)}.

\bibitem{gx} G. X. Ni, A. S. McLeod, Z. Sun \textit{et al.}, Fundamental limits to graphene plasmonics
, \href{https://doi.org/10.1038/s41586-018-0136-9}{Nature {\bf 557}, 530–533 (2018)}.

\bibitem{rig} A. F. Rigosi, H. M. Hill, N. R. Glavin, S. J. Pookpanratana \textit{et al.}, Measuring the dielectric and optical response of millimeter-scale amorphous and hexagonal boron nitride films grown on epitaxial graphene, \href{https://doi.org/10.1088/2053-1583/aa9ea3}{2D Materials {\bf 5}, 011011 (2018)}.

\bibitem{jung} M. Yankowitz, J. Jung, E. Laksono et al., Dynamic band-structure tuning of graphene moiré superlattices with pressure, \href{https://doi.org/10.1038/s41586-018-0107-1}{Nature {\bf 557}, 404–408 (2018)}.

\bibitem{merano} M. Merano, Transverse electric surface mode in atomically thin Boron–Nitride, \href{https://doi.org/10.1364/OL.41.002668}{Opt. Lett. {\bf 41}, 2668 (2016)}.

\bibitem{mak} K. F. Mak, C. Lee, J. Hone, J. Shan, and Tony F. Heinz, Atomically Thin MoS$_{2}$: A New Direct-Gap Semiconductor, \href{https://doi.org/10.1103/PhysRevLett.105.136805}{Phys. Rev. Lett. {\bf 105}, 136805 (2010)}.

\bibitem{toma} A. Tomadin, F. Guinea, and M. Polini, Generation and morphing of plasmons in graphene superlattices, \href{https://doi.org/10.1103/PhysRevB.90.161406}{Phys. Rev. B {\bf 90}, 161406(R) (2014)}.

\end{thebibliography}
\end{document}